\documentclass[12pt]{article}

\usepackage{verbatim}
\usepackage{hyperref}
\usepackage{color}
\usepackage{amsfonts}
\usepackage{amsmath}
\usepackage{amssymb}
\usepackage{latexsym}
\usepackage{textcomp}
\usepackage{lastpage}
\usepackage[pdftex]{graphicx} 
\usepackage[center]{caption2}

\date{}
\begin{document}
\begin{flushright}
\hfill UPR-1283-T\
\end{flushright}

\vspace{20pt}
\begin{center}
{\Large {\bf General relation between Aretakis charge and Newman-Penrose charge}}

\vspace{18pt}
{\large  Mirjam Cveti\v c$^\dagger$ and Alejandro Satz$^{\dagger\dagger}$}

\vspace{8pt}

{\it  $\dagger$ Department of Physics and Astronomy,  University of Pennsylvania, Philadelphia, PA 19104, USA\\ \&  Center for Applied Mathematics and Theoretical Physics, University of Maribor, SI2000 Maribor, Slovenia }

\vspace{5pt}

{\it  $\dagger\dagger$ Sarah Lawrence College, Bronxville, NY 10708, USA \& Department of Physics and Astronomy,  University of Pennsylvania, Philadelphia, PA 19104, USA}

\end{center}
\vspace{26pt}

\begin{abstract}
We reexamine the relation between the Aretakis charge of an extremal black hole spacetime and the Newman-Penrose charge of a weakly asymptotically flat spacetime obtained from the original one through radial inversion and conformal mapping. Building on recent work by Godazgar, Godazgar and Pope, we present an explicit general relation between these quantities showing  how the charge densities are mapped. As a non-trivial  example we provide the computation of both quantities and their explicit relation for the extremal Kerr spacetime.
\end{abstract}

\section{Introduction}

As part of his analyses showing the instability of extreme black hole horizons under perturbations, Aretakis showed that fields propagating on the extreme black hole background have a conserved horizon charge \cite{Aretakis:2011ha, Aretakis:2011hc,Aretakis:2012ei}. This so-called Aretakis charge has been shown to exist in higher dimensions as well as four \cite{Murata:2012ct}, and for massless and massive scalars as well as gauge fields \cite{Lucietti:2012sf,Lucietti:2012xr}. Moreover, although the Aretakis charge involves in its definition only the field and its radial derivative at the horizon, it is only the first in an infinite hierarchy of conserved charges involving higher radial derivatives \cite{Aretakis:2012ei, Lucietti:2012sf}.

Shortly after the discovery of the Aretakis charges, it was shown \cite{Lucietti:2012xr, Bizon:2012we} that for extremal Reissner-Nordstr\"om this charges can be mapped, through a conformal symmetry of the metric under radial inversion, to  Newman-Penrose charges at future asymptotic null infinity. Higher multipoles of this mapping were computed in \cite{Bhattacharjee:2018pqb}. In addition, it has been argued in \cite{Godazgar:2017igz} that the Aretakis charge of any extreme black hole can be put in correspondence,  through a similar procedure, with the Newman-Penrose charge of a dual weakly asymptotic spacetime. However, no other explicit examples than the Reissner-Nordstr\"om were computed

The goal of this paper is to extend and refine the conclusions of \cite{Godazgar:2017igz}. Firstly, we make fully explicit the relation between the Aretakis charge of a general extreme four-dimensional black hole and the Newman-Penrose charge in an associated weakly asymptotically flat spacetime, which is conformally related to the radial inversion of the original spacetime. Secondly, we provide a new nontrivial example by exhibiting this relation for the extremal Kerr black hole. A main conclusion of our work is that when the black hole is not spherically symmetric there is no direct mapping between the two charges but rather a more subtle relation between the charge densities.

In the next section, we establish our framework and summarize the main results of \cite{Godazgar:2017igz}. In Section 3, we present the general relation between the Aretakis and the Newman-Penrose charges, as well as its explicit computation for the extreme Kerr black hole. Section 4 contains a summary and discussion of the results.

\section{Duality between extremal horizon and asymptotic infinity}

In this section we summarize the results of \cite{Godazgar:2017igz} concerning the duality between the Aretakis charge of an extremal black hole and the asymptotic Newman-Penrose charge of a dual conformal metric. 

Consider a 4-dimensional extremal black hole with a Killing horizon. In the vicinity of the horizon one can introduce Gaussian null coordinates $(v,\rho,x^i)$, with $i=(1,2)$. The horizon corresponds to $\rho=0$, and near it the metric takes the form:
\begin{align}\label{GNCPope}
ds^2&=L(x^i)^2\left[-\rho^2\,F(\rho, x^i)dv^2+2dvd\rho\right]\,\nonumber\\&+\,\gamma_{ij}(\rho,x^i)\left(dx^i-\rho\,h^i(\rho,x^i) dv\right)\left(dx^j-\rho\,h^j(\rho,x^i) dv\right)\,,
\end{align}
with $F(\rho=0, x^i)=1$. It is assumed that $h^i(\rho,x^i) = O(1)$ at the horizon, and that $\gamma_{ij}(\rho=0,x^i)$ is a topologically spherical metric.  A massless scalar field $\psi$ on this background has an Aretakis charge defined by
\begin{equation}\label{HAredef}
H_A = \lim_{\rho\to 0}\int d^2x\sqrt{\gamma}\left[ 2\partial_\rho \psi +\frac{1}{2}\frac{\partial_\rho \gamma}{\gamma}\psi\right]\,,
\end{equation}
with $\gamma= \mathrm{det}\, \gamma_{ij}$. The Aretakis charge is conserved on the horizon (i.e. its value is $v$-independent). 

Under the coordinate change 
\begin{equation}
\rho\longrightarrow r = \frac{1}{\rho}\,,
\end{equation}
the metric takes the form
\begin{equation}\label{conf}
ds^2 = \frac{L^2}{r^2}\left[-Fdv^2-2dvdr+r^2h_{ij}\left(dx^i-C^idv\right) \left(dx^j-C^jdv\right) \right]\,,
\end{equation}
where $C^i=h^i/r$ and $h_{ij} = \gamma_{ij}/L^2$. The metric inside the square brackets corresponds to a weakly asymptotically flat spacetime, in which the asymptotic 2-dimensional compact space has metric
\begin{equation}\label{omegaij}
\omega_{ij}(x^i) = \lim_{r\to\infty}h_{ij}(r,x^i) = \frac{\gamma_{ij}(\rho=0,x^i)}{L(x^i)^2}\,.
\end{equation}
We call the weakly asymptotically flat metric in the square brackets of (\ref{conf}) the conformal dual of the initial metric with an extremal black hole.

A massless scalar field $\tilde{\psi}$ propagating in a weakly asymptotically flat spacetime has a conserved (i.e. $v$-independent) charge at future null infinity, namely the Newman-Penrose charge \cite{Newman:1968uj}. Its definition is:
\begin{equation}\label{HNPdef}
H_{NP}=-\lim_{r\to\infty}\int d^2x\,\sqrt{\omega}\left[ 2\partial_r(r\tilde{\psi})+r\partial_r\zeta\,\tilde{\psi} \right]\,.
\end{equation}
Here $\omega=\mathrm{det}\,\omega_{ij}$ and $\zeta$ is defined by the relation:
\begin{equation}\label{zetadef}
 h= \mathrm{det}\,h_{ij} = \omega \zeta^2\,.
 \end{equation}
 Note that weak asymptotic flatness implies $\zeta(r)=1+O(1/r)$ as $r\to\infty$.

In \cite{Godazgar:2017igz} it is shown as well that the solutions to the massless wave equation in the original spacetime and in its conformal dual are related by
\begin{equation}\label{psipsitilde}
\psi(v,\rho,x^i) = \frac{L(x^i)}{r}\tilde{\psi}\left(v,\frac{1}{r},x^i\right)
\end{equation}
It is stated in \cite{Godazgar:2017igz} that this allows us to map the Aretakis charge of the original spacetime to the NP charge of its conformal dual. However the relation is not worked out explicitly except in the particular case of the extremal Reissner-Nordstr\"om solution. In the following section we derive the general explicit relation between both charges.

\section{General relation between Aretakis charge and dual NP charge in 4 dimensions}

To find the explicit general relation between $H_A$ and the conformal dual $H_{NP}$, we write the field in the vicinity of the original spacetime's horizon in an expansion of the form:
\begin{equation}\label{psiexp}
\psi = \psi_0 + \rho \,\psi_1 +O(\rho^2)\,,
\end{equation}
where the $\psi_j$ coefficients depend on $v$ and $x^i$ but not $\rho$.
Using this expansion in (\ref{HAredef}) leads to:
\begin{equation}\label{HArefinal}
H_A = \int d^2x\,\sqrt{\gamma}\left(2\psi_1+\frac{1}{2}\frac{\gamma'}{\gamma}\psi_0\right)\,,
\end{equation}
where both $\gamma$ and its derivative are evaluated at the horizon ($\rho=0$). 

Analogously, writing the asymptotic field in the conformal dual spacetime in an expansion of the form
\begin{equation}\label{psitildeexp}
\tilde{\psi} = \frac{\tilde{\psi}_1}{r}+\frac{\tilde{\psi}_2}{r^2}+O(1/r^3)\,,
\end{equation}
we get when replacing in (\ref{HNPdef}) the result
\begin{equation}
H_{NP} = \int d^2x \,\sqrt{\omega}\left(2\tilde{\psi}_2-r^2\zeta'\tilde{\psi}_1\right)\,,
\end{equation}
where now $\zeta'$ stands for the $r\to\infty$ asymptotic limit of $\partial_r \zeta$.

From (\ref{omegaij}) and (\ref{zetadef}) we have the relation
\begin{equation}
\zeta = L(x^i)^{-2}\frac{\sqrt{\gamma}}{\sqrt{\omega}}\,.
\end{equation}
From this using the chain rule, that  $\rho = 1/r$, and that $\omega = L^{-4}\gamma\big|_{\rho=0}$, we find:
\begin{equation}
\zeta'(r)\Big|_{r\to\infty} = -\frac{1}{2r^2}\frac{\gamma'(\rho)}{\gamma}\Big|_{\rho=0}\,.
\end{equation}
Therefore the NP charge reduces to
\begin{equation}
H_{NP} = \int d^2x \, \sqrt{\omega}\left(2\tilde{\psi}_2+\frac{1}{2}\frac{\gamma'(\rho)}{\gamma}\Big|_{\rho=0}\tilde{\psi}_1 \right)\,.
\end{equation}
Using the relation (\ref{psipsitilde}) to match the expansion coefficients we see that
\begin{equation}
\psi_0 = L(x^i)\,\tilde{\psi}_1\,,\quad\quad \psi_1 = L(x^i)\,\tilde{\psi}_2\,,
\end{equation}
which implies that
\begin{equation}\label{HNPfinal}
H_{NP} = \int d^2x \, \sqrt{\omega}\,L^{-1}(x^i)\left(2\psi_1+\frac{1}{2}\frac{\gamma'}{\gamma}\Big|_{\rho=0}\psi_0 \right)\,.
\end{equation}

Comparing (\ref{HArefinal})  with  (\ref{HNPfinal}) we see that the ``charge densities'' in both expressions (i.e. the quantities that give $H_A$ and $H_{NP}$ when integrated in two dimensions with measures $\sqrt{\gamma}$ and $\sqrt{\omega}$ respectively) are identical up to an additional $L^{-1}(x^i)$ factor in the second one.

Since the relation between both 2d measures is found from (\ref{omegaij}), we can also summarize the relation between both charges as:
\begin{equation}\label{densitymap}
H_A = \int d^2x\,\sqrt{\gamma}\big|_{\rho=0} \,h_A\,,\quad\quad H_{NP} = \int d^2x\, \sqrt{\gamma}\big|_{\rho=0} \,h_{NP}\,;\quad\quad h_A = L(x^i)^3\,h_{NP}\,.
\end{equation}
In other words, the ``charge densities'' to be integrated under the same measure are related by a $L^{3}(x^i)$ factor. 

We see that there is no straightforward mapping between the charges, in the sense that there is no operation that provides the quantity $H_{NP}$ directly from the input $H_A$ or vice versa. Rather, it is the densities to be subjected to angular integration that are related through the factor $L(x^i)$ from the metric (\ref{GNCPope}).

The only case in which the two charges can be directly mapped is when $L(x^i)$ is a constant. An example of this is the extreme Reissner-Nordstr\"om metric, where we have $L=M$ ($M$ being the mass parameter). In this case, our result reduces immediately to the simpler relation $H_A = M^3\,H_{NP}$. This agrees with the result of \cite{Lucietti:2012xr,Godazgar:2017igz}.

\subsection{Example: Extremal Kerr black hole}

As a non-trivial example, in this subsection we compute the Aretakis charge and the conformal dual NP charge for the extremal Kerr black hole.

As discussed e.g.~in \cite{Kunduri:2007vf, Kunduri:2008rs}\footnote{For higher dimensional examples of extremal near horizon geometries, see, e,g., \cite{Chow:2008dp}.}, the near-horizon regime of the extremal Kerr metric is described by Gaussian null coordinate system of the form
\begin{align}
ds^2 &= \left(\frac{1+x^2}{2}\right)\left(-\frac{\tilde{\rho}^2}{2a^2} dv^2+2dvd\tilde{\rho}\right)+a^2\left(\frac{1+x^2}{1-x^2} \right)dx^2\nonumber\\
&+4a^2\left(\frac{1-x^2}{1+x^2} \right)\left(d\phi+\frac{\tilde{\rho}}{2a^2}dv\right)^2\,,
\end{align}
where $x = \cos\theta$.

Comparing with the form of the metric  (\ref{GNCPope}) on which our analysis is based, we see that the radial coordinates $\tilde{\rho}$ and $\rho$ are related by the rescaling $\tilde{\rho} = 2a^2\rho$, and that therefore we have:
\begin{equation}\label{LKerr}
L^2(x)=a^2(1+x^2)\,.
\end{equation}

The angular metric functions are given by
\begin{equation}
 \gamma_{ij}dx^idx^j =  a^2\frac{1+x^2}{1-x^2}dx^2+4a^2\frac{1-x^2}{1+x^2}d\phi^2\,,\label{gammaKerr}
\end{equation}
whereas $h^x=0$ and $h^\phi=1$.
The determinant $\gamma$ is computed from (\ref{gammaKerr}) to be
\begin{equation}
\gamma = 4a^4
\end{equation}

It should be noted that all these metric functions are evaluated at the horizon. For the computation of the Aretakis charge we also need the first-order expansion of $\gamma_{ij}$ away from the horizon. This has been computed in \cite{Li:2015wsa}:

\begin{equation}
\gamma_{xx}^{(1)} = \frac{4a}{1-x^4}\,,\quad\quad \gamma_{\phi\phi}^{(1)} = \frac{16ax^2(1-x^2)}{(1+x^2)^3}\,,\quad\quad \gamma_{x\phi}^{(1)}=\frac{4ax(1-x^2)}{(1+x^2)^2}\,.
\end{equation}
From these and the zeroth-order coefficients given in (\ref{gammaKerr}) it is straightforward to compute:
\begin{equation}\label{gammap}
\frac{\partial_\lambda \gamma}{\gamma}\Big|_{\lambda=0} = \frac{4}{a(1+x^2)}
\end{equation}
The radial expansion parameter $\lambda$ here is not the same as our radial coordinate $\rho$. It follows the form of metric used in \cite{Li:2015wsa}: 
\begin{equation}
ds^2 = \lambda^2 \overline{F}(\lambda,x)dv^2+ 2dv d\lambda+ 2\lambda \overline{h}_i(\lambda,x)dv dx^i +\gamma_{ij}(\lambda,x)dx^idx^j
\end{equation}
This radial coordinate is related to ours by $\lambda=L^2(x)\,\rho$ in the vicinity of the horizon.  Using this, (\ref{LKerr}) and (\ref{gammap}), we can write the Aretakis charge (\ref{HAredef}) of the extremal Kerr black hole as:
\begin{equation}
H_A = 4a^2\int dx\,d\phi \left[ \partial_\rho \psi\big|_{\rho=0} +a\, \psi\big|_{\rho=0}\right]\,.
\end{equation}
 Consequently, the Newman-Penrose charge of the conformal dual spacetime (written in terms of the field and its radial derivative on the original spacetime's horizon) is given by
\begin{equation}
H_{NP} = \frac{4}{a}\int dx\,d\phi\, \frac{1}{(1+x^2)^{3/2}}\,\left[ \partial_\rho \psi\big|_{\rho=0} +a\, \psi\big|_{\rho=0}\right]\,.
\end{equation}
Due to the nontrivial nature of the angular function $L(x)$, there is no direct mapping between both charges as happens in the extreme Reissner-Nordstr\"om case.

We conclude this section with some additional remarks on the relation between the original metric and its conformal dual. The extremal four-dimensional Reissner-Nordstr\"om metric has the property that its conformal dual metric, as defined above from equation (\ref{conf}), is identical to the original one (up to a constant that can be absorbed into a redefinition of the radial coordinate) \cite{couch}. Thus in the special case of extremal Reissner-Nordstrom the Aretakis charge is mapped to the NP charge of the same spacetime. A similar property is enjoyed by extremal 4-charge static black holes in ungauged four-dimensional STU supergravity, in the case where the four charges are pairwise equal \cite{Godazgar:2017igz}. These correspondences were an important part of the motivation of \cite{Godazgar:2017igz} for exploring the relation between Aretakis and NP charges. Therefore, it is interesting to explore whether extremal Kerr satisfies any similar property.

Though we do not have the full expression for the extremal Kerr metric in the horizon-adapted Gaussian null coordinate system, we have the zeroth-order metric at the horizon, which lets us know the zeroth-order asymptotic metric at $r\to\infty$ in the conformal dual spacetime. This metric is:

\begin{equation}
ds^2 = -dv^2-2dv dr + \frac{r^2\,\gamma_{ij}(x)}{L^2(x)}\left(dx^i-\frac{h^i(x)}{r}dv\right)\left(dx^j-\frac{h^j(x)}{r}dv\right)\,,
\end{equation}
with $L$ given in (\ref{LKerr}), $\gamma_{ij}$ given in (\ref{gammaKerr}), and  $h^x=0$, $h^\phi=1$ as before. It is seen that the conformal dual metric is not (asymptotically) equal to the original Kerr metric and in fact is only weakly asymptotically flat (rather than asymptotically flat) due to the nontrivial angular metric. Indeed  after replacing $x = \cos\theta$ we have
\begin{equation}
\frac{\gamma_{ij}dx^idx^j}{L^2} = d\theta^2 + \frac{4\sin^2\theta}{(1+\cos^2\theta)^2}d\phi^2
\end{equation}
instead of $d\theta^2+\sin^\theta d\phi^2$ as should be the case in an asymptotically flat spacetime. The simple relation between the extremal black hole spacetime and its conformal dual present for Reissner-Nordstr\"om spacetime is therefore not generalized to rotating black holes.

\section{Discussion}

The primary goal of the present paper is to extend the results of \cite{Godazgar:2017igz} by providing an explicit expression linking the Aretakis charge of an extremal black hole spacetime with the Newman-Penrose charge of the conformal dual spacetime. By the latter we mean the conformal metric to the $\rho\longrightarrow 1/r$ inverted spacetime, as explained above (eq. \ref{conf}). We saw that, aside from spherically symmetric cases such as the extremal Reissner-Nordstr\"om black hole, in general there is no direct mapping between the charges. Rather, the mapping is at the level of the densities to be integrated over the horizon and over asymptotic infinity, as exhibited above through  (\ref{HArefinal}) and (\ref{HNPfinal}), or more explicitly in (\ref{densitymap}).

We have also provided an explicit computation of both charges in the extremal Kerr spacetime (and its conformal dual). This is noteworthy as a nontrivial example where the metric is not spherically symmetric. The calculation was facilitated by the previous construction in  \cite{Li:2015wsa} of the first-order expansion for the angular near-horizon metric $\gamma_{ij}$.

The results presented in this paper should be a stepping stone towards further exploration of the near horizon-asymptotic infinity duality and its consequences. One area in which additional investigations would be especially fruitful would be the extension of the results presented to higher dimensions, which was already hinted in \cite{Godazgar:2017igz} but could be made explicit along the lines of the present paper. In higher dimensions the correspondence between $H_A$ and $H_{NP}$ can still be derived, although the  dual spacetime on which $H_{NP}$ is defined is less likely to admit of a physical interpretation because the definition of weak asymptotic flatness required for the derivation ($\zeta= 1 +O(1/r)$) is further away from ordinary asymptotic flatness than it is in four dimensions. Exploring the issue in depth could possibly help clarify as well the physical meaning of this correspondence in four dimensions.

Another area of application which already received preliminary discussion in \cite{Godazgar:2017igz} is extremal 4d black holes in STU supergravity \cite{Cvetic:1995uj,Cvetic:1995bj, Cvetic:1996kv, Chow:2014cca} and their  generalizations, including a study of  the Aretakis charge and its dualities for the extremal version of their ``subtracted geometry'' limit \cite{Cvetic:2011hp, Cvetic:2011dn,Cvetic:2014sxa}.

\section*{ Acknowledgements} 

\begin{thebibliography}{10}


\bibitem{Aretakis:2011ha} 
  S.~Aretakis,
  ``Stability and Instability of Extreme Reissner-Nordstr\'om Black Hole Spacetimes for Linear Scalar Perturbations I,''
  Commun.\ Math.\ Phys.\  {\bf 307}, 17 (2011)
  [arXiv:1110.2007 [gr-qc]].


\bibitem{Aretakis:2011hc} 
  S.~Aretakis,
  ``Stability and Instability of Extreme Reissner-Nordstrom Black Hole Spacetimes for Linear Scalar Perturbations II,''
  Annales Henri Poincare {\bf 12}, 1491 (2011)
  [arXiv:1110.2009 [gr-qc]].


\bibitem{Aretakis:2012ei} 
  S.~Aretakis,
  ``Horizon Instability of Extremal Black Holes,''
  Adv.\ Theor.\ Math.\ Phys.\  {\bf 19}, 507 (2015)
  [arXiv:1206.6598 [gr-qc]].

\bibitem{Murata:2012ct} 
  K.~Murata,
  Class.\ Quant.\ Grav.\  {\bf 30}, 075002 (2013)
  [arXiv:1211.6903 [gr-qc]].

\bibitem{Lucietti:2012sf} 
  J.~Lucietti and H.~S.~Reall,
  ``Gravitational instability of an extreme Kerr black hole,''
  Phys.\ Rev.\ D {\bf 86}, 104030 (2012)
  [arXiv:1208.1437 [gr-qc]].
  
\bibitem{Lucietti:2012xr} 
  J.~Lucietti, K.~Murata, H.~S.~Reall and N.~Tanahashi,
  ``On the horizon instability of an extreme Reissner-Nordstr\'om black hole,''
  JHEP {\bf 1303}, 035 (2013)
  [arXiv:1212.2557 [gr-qc]].
  
\bibitem{Bizon:2012we} 
  P.~Bizon and H.~Friedrich,
  ``A remark about wave equations on the extreme Reissner-Nordstr\'om black hole exterior,''
  Class.\ Quant.\ Grav.\  {\bf 30}, 065001 (2013)
  [arXiv:1212.0729 [gr-qc]].
  
\bibitem{Bhattacharjee:2018pqb} 
  S.~Bhattacharjee, B.~Chakrabarty, D.~D.~K.~Chow, P.~Paul and A.~Virmani,
  Class.\ Quant.\ Grav.\  {\bf 35}, no. 20, 205002 (2018)
  [arXiv:1805.10655 [gr-qc]].

\bibitem{Godazgar:2017igz} 
  H.~Godazgar, M.~Godazgar and C.~N.~Pope,
  ``Aretakis Charges and Asymptotic Null Infinity,''
  Phys.\ Rev.\ D {\bf 96}, no. 8, 084055 (2017)
  [arXiv:1707.09804 [hep-th]].
  
\bibitem{Newman:1968uj} 
  E.~T.~Newman and R.~Penrose,
  ``New conservation laws for zero rest-mass fields in asymptotically flat space-time,''
  Proc.\ Roy.\ Soc.\ Lond.\ A {\bf 305}, 175 (1968).
  
\bibitem{Kunduri:2007vf} 
  H.~K.~Kunduri, J.~Lucietti and H.~S.~Reall,
  ``Near-horizon symmetries of extremal black holes,''
  Class.\ Quant.\ Grav.\  {\bf 24}, 4169 (2007)
  [arXiv:0705.4214 [hep-th]].
  
\bibitem{Kunduri:2008rs} 
  H.~K.~Kunduri and J.~Lucietti,
  ``A Classification of near-horizon geometries of extremal vacuum black holes,''
  J.\ Math.\ Phys.\  {\bf 50}, 082502 (2009)
  [arXiv:0806.2051 [hep-th]].
  
\bibitem{Chow:2008dp} 
  D.~D.~K.~Chow, M.~Cveti\v c, H.~L{\"u} and C.~N.~Pope,
 ``Extremal Black Hole/CFT Correspondence in (Gauged) Supergravities,''
  Phys.\ Rev.\ D {\bf 79}, 084018 (2009)
  [arXiv:0812.2918 [hep-th]].
\bibitem{Li:2015wsa} 
  C.~Li and J.~Lucietti,
  ``Transverse deformations of extreme horizons,''
  Class.\ Quant.\ Grav.\  {\bf 33}, no. 7, 075015 (2016)
  [arXiv:1509.03469 [gr-qc]].
  
  \bibitem{couch}
  W.~E.~Couch and R.~J.~Torrence, 
  ``Conformal invariance under spatial inversion of
extreme Reissner-Nordstr\"om black holes,''
Gen.\ Rel.\ \& Grav.\ {\bf 16} no. 8, 789 (1984). 

\bibitem{Cvetic:1995uj} 
  M.~Cveti\v c and D.~Youm,
  ``Dyonic BPS saturated black holes of heterotic string on a six torus,''
  Phys.\ Rev.\ D {\bf 53}, 584 (1996)
  [hep-th/9507090].
\
\bibitem{Cvetic:1995bj} 
  M.~Cveti\v c and A.~A.~Tseytlin,
  ``Solitonic strings and BPS saturated dyonic black holes,''
  Phys.\ Rev.\ D {\bf 53}, 5619 (1996)
  Erratum: [Phys.\ Rev.\ D {\bf 55}, 3907 (1997)]
  [hep-th/9512031].


\bibitem{Cvetic:1996kv}
  M.~Cveti\v c and D.~Youm,
  ``Entropy of nonextreme charged rotating black holes in string theory,''
  Phys.\ Rev.\ D {\bf 54}, 2612 (1996)
  [hep-th/9603147].

\bibitem{Chow:2014cca}
  D.~D.~K.~Chow and G.~Comp{\` e}re,
  ``Black holes in N=8 supergravity from SO(4,4) hidden symmetries,''
  Phys.\ Rev.\ D {\bf 90}, no. 2, 025029 (2014)
  [arXiv:1404.2602 [hep-th]].
  


\bibitem{Cvetic:2011hp} 
  M.~Cveti\v c and F.~Larsen,
``Conformal Symmetry for General Black Holes,''
  JHEP {\bf 1202}, 122 (2012)
  [arXiv:1106.3341 [hep-th]].

\bibitem{Cvetic:2011dn} 
  M.~Cveti\v c and F.~Larsen,
``Conformal Symmetry for Black Holes in Four Dimensions,''
  JHEP {\bf 1209}, 076 (2012)
  [arXiv:1112.4846 [hep-th]].

\bibitem{Cvetic:2014sxa} 
  M.~Cveti\v c and F.~Larsen,
``Black Holes with Intrinsic Spin,''
  JHEP {\bf 1411}, 033 (2014)
  [arXiv:1406.4536 [hep-th]].

\end{thebibliography}

We would like to thank Hadi Godazgar,  Mahdi Godazgar and Christopher Pope for valuable discussions, especially on the higher dimensional generalizations of the current work.  M.C. would like to thank  the
Mitchell Family Foundation for hospitality at the Brinsop Court workshop. The research of M.C.~and A.S. is supported in part by the University of Pennsylvania Research Foundation (M.C. and A.S.),   the DOE Award DE-SC0013528 (M.C.),  the  Fay  R.  and  Eugene  L.  Langberg  Endowed  Chair  (M.C.)  and  the
Slovenian Research Agency (M.C.).

\end{document}